\documentclass[amsmath,10pt,twocolumn, superscriptaddress, footinbib,pre]
{revtex4-1}
\usepackage{amsmath,amsfonts,bm}
\usepackage{graphicx}
\usepackage{color}
\usepackage{subfigure}

\newcommand{\matB}[1]{\left( \begin{smallmatrix} \frac{#1+1+x}{x^{1-\lambda}} 
&& -x^{2\lambda-1}\\1 && 0\end{smallmatrix}\right)}
\newcommand{\matA}[1]{\left( \begin{smallmatrix} \frac{#1+h+x}{h} 
&& -\frac{x^{\lambda}}{h}\\ 1 && 0\end{smallmatrix}\right)}
\newcommand{\matAA}[1]{\left( \begin{smallmatrix} \frac{#1+1+h}{x^{1-\lambda}}
&& -h x^{\lambda-1}\\1 && 0\end{smallmatrix}\right)}

\begin{document}
\title{Dynamic phase transitions in simple driven kinetic networks}
\author{Suriyanarayanan Vaikuntanathan}
\affiliation{Material Sciences Division, Lawrence Berkeley National Lab, 
Berkeley, CA 94720}
\author{Todd R. Gingrich}
\affiliation{Department of Chemistry, University of California, Berkeley, CA 
94720}
\author{Phillip L. Geissler}
\affiliation{Material Sciences Division, Lawrence Berkeley National Lab, 
Berkeley, CA 94720}
\affiliation{Department of Chemistry, University of California, Berkeley, 
CA 94720}
\affiliation{Chemical Sciences Division, Lawrence Berkeley National Lab, 
Berkeley, CA 94720}

\begin{abstract}
We analyze the probability distribution for entropy production rates of 
trajectories evolving on a class of out-of-equilibrium kinetic networks.  
These networks can serve as simple models for driven dynamical systems, 
where energy 
fluxes typically result in non-equilibrium dynamics.
By analyzing the fluctuations in the entropy production, we demonstrate the 
emergence, in a large system size limit, of a dynamic phase transition between 
two distinct dynamical regimes. 
\end{abstract}
\maketitle 

\begin{figure}[tbhp]         
\centering
\subfigure{
\includegraphics[width=0.98\linewidth]{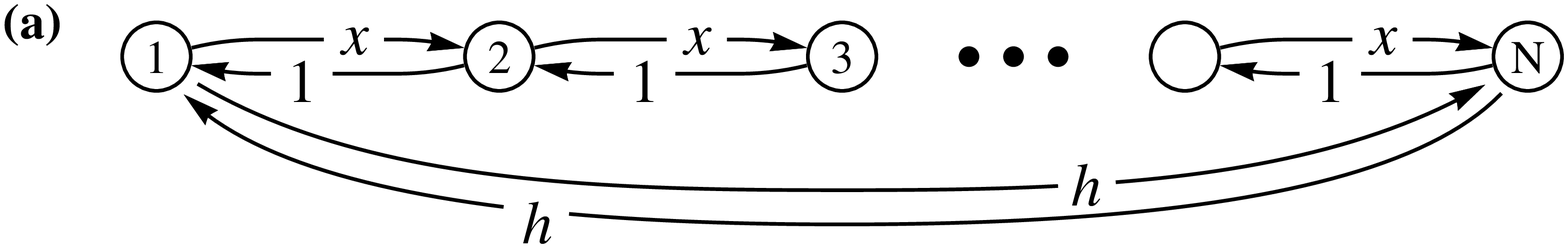} 
\label{fig:ringnetwork}
}
\subfigure{
\includegraphics[width=0.98\linewidth]{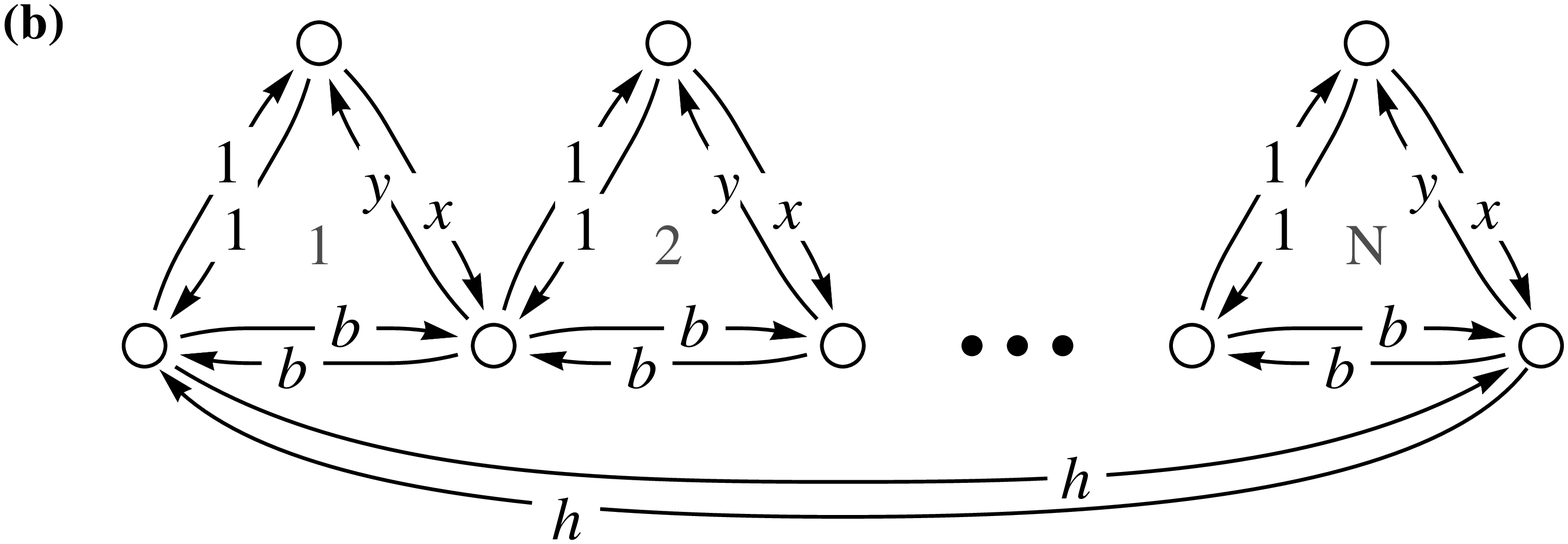}
\label{fig:trianglenetwork}
}
\caption{Diagrams of simple driven kinetic networks studied here.  
Arrows connecting a pair of vertices indicate 
Poisson-distributed transitions from one state to the other, labeled by the 
corresponding rate constants.}
\label{fig:network}
\end{figure}

The study of fluctuation phenomena is one of the central endeavors of 
non-equilibrium statistical mechanics.  Analysis of fluctuations in 
non-equilibrium processes have, for example, led to the discovery of the 
fluctuation theorems, which have helped elucidate how macroscopic notions of 
irreversibility emerge from microscopic laws~\cite{Crooks2000,Jarzynski1997,
Jarzynski2011}.  
More recently, theoretical and numerical analysis of the statistics of rare 
fluctuations in driven lattice gas models~\cite{Bodineau2004,Bodineau2005}, 
exclusion processes~\cite{Espigares2013}, zero-range processes~\cite{
Harris2005}, 1D models of transport~\cite{Hurtado2011v2}, and models of 
glass formers~\cite{Garrahan2007,Speck2012v2} have revealed the presence of 
coexisting ensembles of trajectories and so-called dynamic phase transitions 
between them~\cite{Bodineau2004,Bodineau2005,Hurtado2011v2,Bunin2013}.
In this paper, we analyze the statistics of rare fluctuations in entropy 
production rates for certain model non-equilibrium, or driven, kinetic 
networks (see Fig.~\ref{fig:network}).
While this Markovian system, with effectively one-particle dynamics, lacks 
much of the complexity of previously studied driven systems~\cite{Bodineau2004,
Bodineau2005, Espigares2013, Hurtado2011v2,Harris2005}, we show 
\textemdash \ numerically and analytically \textemdash \ the presence of two 
dynamical phases, each with a characteristic entropy production rate. 
This demonstration shows that singularities in trajectory space can in fact arise 
even in very simple driven kinetic networks with a single degree of freedom.
Driven kinetic networks of this general flavor are used to model a variety of physical, 
chemical, and biological systems including molecular motor 
dynamics~\cite{Kolomeisky2007,Fisher1999}; cellular feedback, control, and 
regulation~\cite{Tu2008}; and kinetic proofreading 
mechanisms~\cite{Hopfield1974,Murugan2012v2}. 
Physically, the dynamic phase transition serves to enhance the probability of 
observing large fluctuations in the dynamical behavior of these hopping processes.
%In addition to elucidating how multiple dynamic regimes can emerge in these 
%networks, this work is of particular interest for its applications 
%to biological systems.  Driven kinetic networks are a ubiquitous tool for 
%modeling out-of-equilibrium dynamics, appearing in models of actin, kinesin, 
%and microtuble growth~\cite{Dogterom1993,Holy1994,Antal2007v2,Wolynes2006}; 
%F0-F1 ATPase and other molecular machines~\cite{Kolomeisky2007}; cellular 
%feedback, control, and regulation~\cite{Tu2008}; and kinetic proofreading 
%mechanisms~\cite{Hopfield1974,Bennett1979,Murugan2012v2} for high-fidelity 
%DNA replication.

%In particular, viewing our networks as very crude models for the cell cycle, this work
%suggests that having heterogeneous rates in the chemical processes underlying
%cell replication should lead to a dynamical heterogeneity that may be related
%to the problem of bacterial persisters~\cite{Balaban2004}.

We study dynamical fluctuations of a system evolving on cyclical or periodic driven kinetic networks with some heterogeneity in 
the transition rates. 
We consider two types of cyclic networks, which we hereafter refer to as the 
ring network and the triangle network.
The ring network connects $N$ states in a circle with transition rates $x$ in 
the clockwise direction and $1$ in the reverse direction.
The network has translational symmetry, but we also construct a variation of 
the ring network with that symmetry broken by a link we call the heterogeneous
 link, or $h$-link.
As shown in Fig.~\ref{fig:ringnetwork}, this link connects states $1$ and $N$ 
with rate $h$ in each direction.
The triangle networks are similar in structure but consist of triangular motifs
as depicted in Fig.~\ref{fig:trianglenetwork}. 
Each triangular motif has one asymmetric link with rates $x \neq y$ resulting 
in cycling currents on average.
Triangular subunits offer both driven and undriven paths between sites on 
their horizontal edge~\footnote{Such motifs were introduced, for example, in some of 
the earliest models of kinetic proofreading~\cite{Hopfield1974,Bennett1979}}.
The important motivation for considering this decoration is to
establish a generality that includes cases in which detailed balance
is violated locally as well as globally.

To demonstrate the presence of multiple dynamical phases, we focus on 
fluctuations in the entropy production rate, $\sigma$, in the large $N$ limit.
This rate is of particular physical interest since it is a measure of the 
power provided by external sources to drive the system through the network in a manner that violates 
detailed balance.
A trajectory on our network corresponds to a series of Poissonian transitions,
or ``hops,'' along links with forward and reverse rate constants $k_f$ and 
$k_r$, respectively.
The total entropy, in units of $k_{\rm B}$, produced as the system evolves 
along a particular trajectory is given by~\cite{Seifert2012}
\begin{equation}
\omega = \sum_{\text{hops}} \ln \frac{k_f}{k_r}.
\label{eq:entropyproductiondef}
\end{equation}

\begin{figure}[tbhp]         
\includegraphics[width=0.85\linewidth]{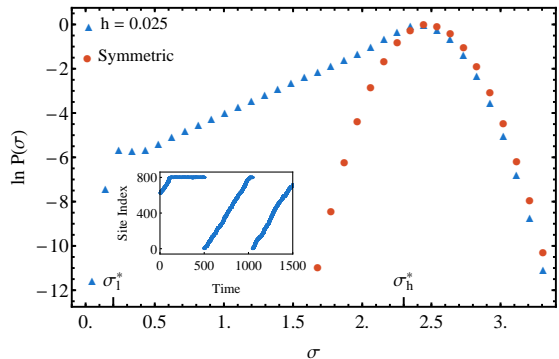}
\caption{
Entropy production distribution for the translationally symmetric triangle 
networks (red) and triangle networks with the heterogeneous link (see 
Fig.~\ref{fig:trianglenetwork}) (blue).  
For each network, $10^7$ independent trajectories of length $\tau = 250$ were 
generated with $x = 20$, $y = 1$, $b = 0.1$, $h = 0.025$, and $N = 400$.
The inset shows a trajectory illustrating the two dynamical regimes.
Sites are numbered clockwise in a zig-zag fashion with the $h$-link 
connecting sites $1$ and $800$.}
\label{fig:simulationdata}
\end{figure}

We first describe numerically sampled steady state trajectories, of length 
$\tau$, and the resulting probability distribution $P(\sigma)$ of the entropy 
production rate $\sigma\equiv \omega/\tau$.
In Fig.~\ref{fig:simulationdata}, we plot $\ln{P(\sigma)}$ for both the 
translationally symmetric triangle network and the triangle network with the 
heterogeneous link, as measured by kinetic Monte Carlo simulations~\cite{
Gillespie1976}. 
For entropy production rates above a critical value, 
$\sigma_\text{h}^*\approx 2.3$, the probability density $P(\sigma)$ of the 
translationally symmetric and heterogeneous networks are almost identical.  
In fact, the most probable entropy production rate is largely unaffected by 
the broken symmetry.
However, for entropy production rates lower than $\sigma_\text{h}^*$, 
$P(\sigma)$ of the heterogeneous network differs from its translationally 
symmetric counterpart by a ``fat tail" which indicates the presence of a 
second distinct dynamical phase, whose entropy production rate is centered 
around $\sigma_\text{l}^* \approx 0.2$.

\begin{figure}[tbhp]         
\includegraphics[width=0.85\linewidth]{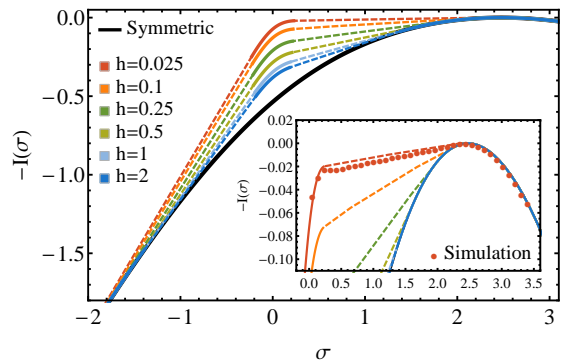}
\caption{
Large deviation function $I(\sigma)$ for entropy production rate of the 
network in Fig.~\ref{fig:trianglenetwork} with $x=20$, $y=1$, $b=0.1$, and 
several different values of $h$.
The rate function envelope was determined by the LF transform of 
$\psi_\omega(\lambda)$, whose two singularities require the construction of 
tie lines (dashed). 
Kinetic Monte Carlo simulation results ($10^7$ trajectories with $h=0.025, 
N=400,$ and $\tau = 250$) are shown as red circles in the inset.}
\label{fig:ratefunction}
\end{figure}

To clarify the nature of these dynamical phases, we harvested a long 
trajectory on the heterogeneous network with a value of $\sigma$ between
$\sigma_\text{l}^*$ and $\sigma_\text{h}^*$.  
This trajectory, shown as an inset in Fig.~\ref{fig:simulationdata} 
illustrates a switching between two types of behaviors.
Initially the trajectory is localized and generates entropy at a rate of 
roughly $\sigma_\text{l}^*$ as it cycles around triangular motifs near the 
$h$-link.
Eventually the trajectory escapes through the $h$-link and rapidly cycles
around the entire network while generating entropy at a rate of approximately
$\sigma_\text{h}^*$.
For very small values of $h$, the behavior is explained simply: a bottleneck 
that hinders repeated cycling through the network should clearly produce some
degree of transient stalling.
In Fig.~\ref{fig:ratefunction}, however, we illustrate that the fat tail in
the entropy production distribution persists even when $h$ is not small
relative to the other rates.
Futhermore, the simple bottleneck explanation does not transparently reveal
the emergence of a true phase transition in trajectory space.

To understand the dynamical fluctuations more generally it is productive to
view the $h$-link as an impurity among otherwise translationally symmetric
units. 
This impurity breaks symmetry and enables trajectories to be split into two
distinct classes.
One class looks like the ordinary random walkers on the translationally
symmetric network with the typical trajectory not stalling around the heterogeneity,
 but the second class localizes around the impurity even when $h>x$.

In order to gain analytical insight into the nature of the dynamic phase transition, 
we consider the scaled cumulant generating 
function~\cite{Touchette2009, Lebowitz1999} of the entropy production, 
\begin{equation}
\psi_\omega(\lambda) = \lim_{\tau \rightarrow \infty} \frac{1}{\tau} \ln 
\left<e^{-\lambda \omega}\right>,
\label{eq:cgf}
\end{equation}
where the expectation value is taken over trajectories initialized in the 
steady state distribution.  In the limit of large $\tau$, the probability of 
observing a particular value of entropy production obeys a large deviation 
principle~\cite{Lebowitz1999},
\begin{equation}
P(\sigma) \approx e^{-\tau I(\sigma)}\,
\label{eq:largedeviationprinciple}
\end{equation}
where $I(\sigma)$ is the large deviation rate function.  
For finite $\tau$, $P(\sigma)$ can in principle be determined by sampling 
trajectories as described above.
Deviations from $\sigma = \langle \sigma \rangle$ become extremely rare, 
however, as $\tau$ grows, so that the limiting form of $I(\sigma)$ is 
impractical to determine by straightforward simulation.
Alternatively the convex envelope of $I(\sigma)$ can be computed as the 
Legendre-Frenchel (LF) transform of $\psi_\omega(\lambda)$~\cite{Touchette2009,
Lebowitz1999}.
Following the general framework laid out by Lebowitz and Spohn~\cite{
Lebowitz1999}, we calculate $\psi_\omega(\lambda)$ as the maximum eigenvalue 
of a matrix operator, $\mathbb{W}_\omega(\lambda)$, which is simply related 
to $\mathbb{W}$, the transition matrix for the kinetic network~\cite{
Lebowitz1999}.  
Specifically the matrix elements of the so-called tilted operator are given by
\begin{equation}
\mathbb{W}_\omega(\lambda)_{ij} = \left(1 - \delta_{ij}\right)
\mathbb{W}_{ij}^{1-\lambda} \mathbb{W}_{ji}^{\lambda} + \delta_{ij} 
\mathbb{W}_{ij}.
\label{eq:Woperatordef}
\end{equation}
By solving for the eigenspectrum of $\mathbb{W}_\omega(\lambda)$ we obtain 
$\psi_\omega(\lambda)$ and therefore the envelope of $I(\sigma)$ via the LF 
transform.  
Furthermore, the form of the maximal eigenvector reflects the character of the
dominant trajectories, as will be addressed in subsequent work.

\begin{figure}[tbhp]         
\includegraphics[width=0.85\linewidth]{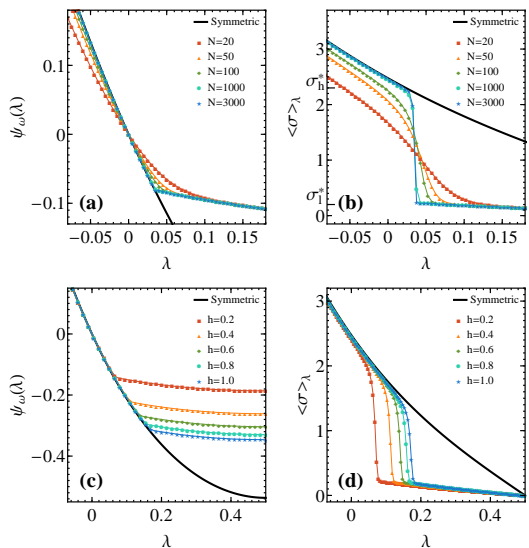}
\caption{Scaled cumulant generating functions and average 
entropy production rates for the triangle networks, depicted in 
Fig.~\ref{fig:trianglenetwork}.  
All results are shown for rates $x = 20, \ y = 1,$ and $b = 0.1$.  
Black solid curves show the exact behavior of the symmetric variant for 
comparison.
Plots (a) and (b) show results for a variety of network sizes with $h = 0.1$ 
for the symmetry-breaking link, suggesting a singularity in the large $N$ limit.
Plots (c) and (d) show results for $N = 200$ and several values of $h$.}
\label{fig:pushdown}
\end{figure}

Consistent with the two-phase behavior suggested by our simulation results, 
numerical eigenvalue calculations for the heterogeneous networks indicate that
$\psi_\omega(\lambda)$ develops a cusp at some critical value $\lambda^*$ in 
the large $N$ limit.  
A second cusp at $1-\lambda^*$ is necessitated by the symmetry of 
$\psi_\omega(\lambda)$ about $\lambda=1/2$ \cite{Lebowitz1999}.  
The emergence of the cusps as $N$ is increased is demonstrated by plots of 
$\psi_\omega(\lambda)$ as a function of $\lambda$ for the triangle network at 
multiple values of $N$ in Fig~\ref{fig:pushdown}(a).  
Plots of $\left<\sigma\right>_\lambda \equiv -{\partial \psi_\omega(\lambda)}
/{\partial \lambda}$ in Fig.~\ref{fig:pushdown}(b) highlight the sharp 
decrease in the first derivative of the cumulant generating function near 
$\lambda=\lambda^*$, which trends towards a discontinuous jump as $N$ 
increases.
The discontinuous change in $\left<\sigma\right>_\lambda$ signals a dynamic 
phase transition, wherein trajectories switch between the characteristic 
entropy production rates, $\sigma_\text{h}^*$ and $\sigma_\text{l}^*$, 
for the two coexisting dynamical phases, in response to a small change in 
$\lambda$.  
We also evaluated $\psi_\omega(\lambda)$ and $\left<\sigma\right>_\lambda$ at 
multiple values of $h$ with fixed $N$.  
These results, collected in Figs.~\ref{fig:pushdown}(c) and (d), show that the
general features described above are present for all the values of $h$ 
considered, and that $h$ serves to tune the critical value of $\lambda$.  
While it is not straightforward to physically bias the $\lambda$-field (it 
couples to a time-nonlocal order parameter), the singularity in 
$\psi_\omega(\lambda)$ provides significant information about the large 
fluctuations in the natural dynamics.

Indeed, the cusp in $\psi_\omega(\lambda)$ when $N\to \infty$ implies that the 
region of the large deviation function between the entropy rates 
$\sigma_\text{h}^*$ and $\sigma_\text{l}^*$ is connected by a Maxwell 
construction (or a tie-line) with a slope $\lambda^*$.
The LF transform of $\psi_\omega(\lambda)$ only provides the convex envelope 
of $I(\sigma)$, but in the limit of large $\tau$, $I(\sigma)$ must converge to
 that envelope~\cite{Touchette2013}. 
Further, as illustrated in Fig.~\ref{fig:pushdown}, $\psi_\omega(\lambda)$ 
converges to the value of its translationally symmetric variant as 
$N\to \infty$ for $\lambda<\lambda^*$. Hence, we expect the large deviation 
rate function to equal that of the translationally symmetric network for 
$\sigma>\sigma_\text{h}^*$, which is illustrated in 
Fig.~\ref{fig:ratefunction}. 
In particular, the inset shows that results from kinetic Monte Carlo 
simulations are consistent with the convex envelope.

These observations can be further clarified and made more rigorous by an 
analytical treatment of the ring and triangle networks. We first focus on the 
ring networks. The simplicity of these networks allows us to transparently 
trace the origin of the cusp and the physical nature of the second dynamical 
phase back to the broken translational symmetry.
Without loss of generality we take $x>1$ and $\lambda<0.5$. 
In the translationally symmetric variant of the ring network, the matrix 
$\mathbb{W}_\omega(\lambda)$ can be diagonalized by a discrete Fourier 
transform to give eigenvalues of the form
\begin{equation}
\label{eq:eigenperiod}
\phi^\text{ts}(\lambda,q) =e^{-2\pi i q/N} x^{1-\lambda}+e^{2\pi i q/N} x^{\lambda}-1-x,
\end{equation}
where $0<q<N-1$.  The superscript ``ts" serves to emphasize the fact that 
Eq.~\ref{eq:eigenperiod} applies only to the translationally symmetric network.
We note that the largest eigenvalue is given when $q=0$, such that 
$\psi^\text{ts}_\omega(\lambda) = \phi^\text{ts}(\lambda, 0)$.
Thus, for the translationally symmetric network, the scaled cumulant 
generating function is smooth and the rate function resulting from the LF 
transform of $\psi_\omega(\lambda)$ is a simple convex function peaked around 
the average entropy production rate. With no broken symmetry there is only 
one dynamical state.

The second phase emerges in networks with the $h$ link. 
We write the right eigenvector corresponding to the largest eigenvalue of 
$\mathbb{W}_\omega(\lambda)$ as $(f_1, f_2, \hdots f_N)$.
Because the matrix is tridiagonal, we can recast the eigenvalue problem as
\begin{equation}
\begin{pmatrix}
f_{1}\\f_{2}
\end{pmatrix} = B^{N-2} A_2 A_1 \begin{pmatrix}f_{1}\\f_{2}\end{pmatrix},
\label{eq:BAAequality}
\end{equation}
where 
\begin{equation}
B = \matB{\psi_\omega(\lambda)}\,,
\end{equation}
and
\begin{equation}
A_1 = \matA{\psi_\omega(\lambda)} \quad A_2 = \matAA{\psi_\omega(\lambda)}.
\end{equation}
The eigenvalues of  $\mathbb{W}_\omega(\lambda)$, in particular the largest 
eigenvalue $\psi_\omega(\lambda)$, can be obtained using 
Eq.~\ref{eq:BAAequality}, which requires that the matrix $B^{N-2} A_2 A_1$ 
have an eigenvalue $1$. The corresponding eigenvector gives 
$\left(f_1,f_2\right)$. 
The other elements of the maximal eigenvector can be obtained using the 
transfer matrices $B$, $A_1$, and $A_2$. 
Specifically, for nodes $n-1$, $n$, and $n+1$ not touching the $h$-link, the 
matrix $B$ maps the eigenvector magnitudes $\left( f_n, f_{n+1}\right)$ onto 
$\left( f_{n-1},f_n\right)$. The matrices $A_1$ and $A_2$ handle similar 
mappings on either side of the heterogeneous link.

In the large $N$ limit, the system with the $h$-link can be solved using a 
perturbative expansion around the solution of the translationally symmetric 
network.
Specifically, we Taylor expand in powers of $1/N$,
\begin{equation}
\psi_\omega(\lambda) = \psi^\text{ts}_\omega(\lambda) + 
\frac{\gamma(x^\lambda-x^{1-\lambda})}{N} + \mathcal{O}\left(
\frac{1}{N^2}\right), 
\label{eq:pertubativeexpansion}
\end{equation}
where we have chosen to express the first-order coefficient in this particular form to 
simplify subsequent algebra.  
Writing $\psi_\omega(\lambda)$ in the form of Eq.~\eqref{eq:pertubativeexpansion} allows us to express $B$ as
\begin{equation}
B = B[\psi^\text{ts}_\omega(\lambda)] + \begin{pmatrix} \frac{\gamma(x^\lambda - x^{1-\lambda})}{Nx^{1-\lambda}} & 0\\ 0 & 0\end{pmatrix} + \mathcal{O}\left(\frac{1}{N^2}\right)
\label{eq:Bexpansion}
\end{equation}
The eigenvalues of $B$ are handled perturbatively.  To first order in $1/N$, the eigenvalues of $B$ can be expressed as
\begin{align}
\nonumber k_1 &= 1 - \frac{\gamma}{N} + \mathcal{O}\left(\frac{1}{N^2}\right) = e^{-\gamma / N} + \mathcal{O}\left(\frac{1}{N^2}\right)\\
k_2 &= x^{2\lambda - 1} + \frac{x^{2\lambda -1}\gamma}{N} + \mathcal{O}\left(\frac{1}{N^2}\right) = x^{2\lambda - 1}e^{\gamma / N} + \mathcal{O}\left(\frac{1}{N^2}\right)
\label{eq:perturbedeigenvalues}
\end{align}

For $x>1$ and $\lambda < 1/2$ and provided $\gamma$ is finite, $k_1 > 1$ and $k_2 < 1$.  
Hence when $B$ is raised to a very large power, only the larger eigenvalue, $k_1$, survives.
\begin{widetext}
\label{eq:largeNoperator}
\begin{equation}
\label{eq:perturb2}
\lim_{N\rightarrow \infty} B^{N-2} = \lim_{N\rightarrow \infty} \frac{e^{-\gamma} + \mathcal{O}(1/N)}{1 - x^{2\lambda - 1}}\begin{pmatrix} 1 & -x^{2 \lambda - 1}\\1 & -x^{2 \lambda - 1}\end{pmatrix} =  \frac{e^{-\gamma}}{1 - x^{2\lambda - 1}}\begin{pmatrix} 1 & -x^{2 \lambda - 1}\\1 & -x^{2 \lambda - 1}\end{pmatrix}
\end{equation}
\end{widetext}

\begin{figure}[thbp]
\centering
\includegraphics[width=1.0\linewidth]{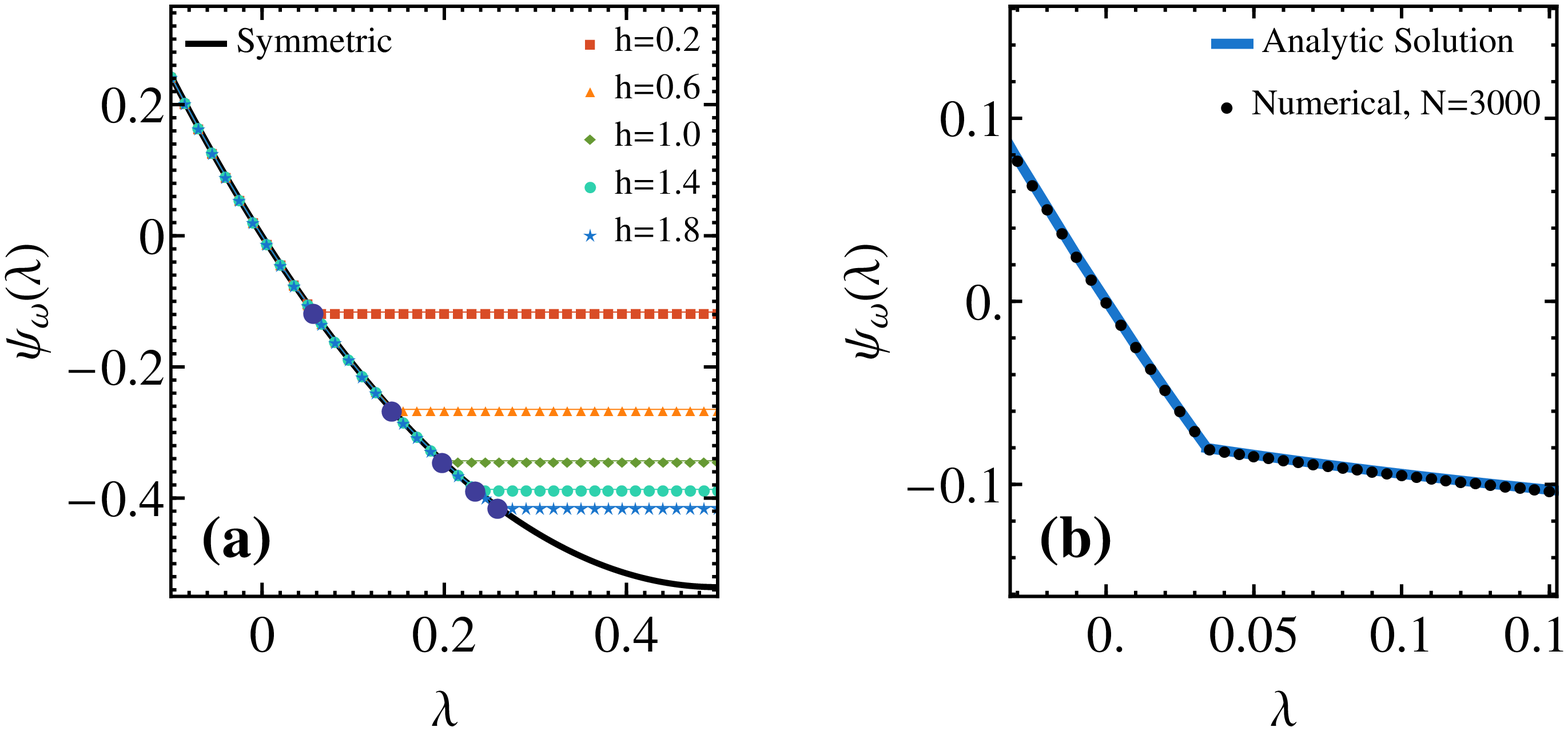}
\caption{ (a) Scaled cumulant generating function for the ring network with varying $h$.  
The value of $\lambda^*$ predicted by our theory is plotted as a large blue point in plot, 
which coincides with the large $N$ limit of the numerical calculations.
(b)Singularity in the scaled cumulant generating 
function for the triangle network with rate constants $x=20, \ y=1,\ b=0.1,$ and $h=0.1$.  
}
\label{fig:crossover}
\end{figure}

To obtain the value of the cumulant generating function we use the condition that $B^{N-2} A_2 A_1$ has an eigenvalue equal to $1$.
From Eq.~\ref{eq:perturb2},
\begin{widetext}
\begin{align}
\label{eq:perturb3}
\nonumber \lim_{N \to \infty} B^{N-2} A_2 A_1 &= \frac{e^{-\gamma}}{1 - x^{2\lambda - 1}}\begin{pmatrix} 1 & -x^{2 \lambda - 1}\\1 & -x^{2 \lambda - 1}\end{pmatrix} \begin{pmatrix} \frac{x^{1-\lambda} + x^{\lambda} + h - x}{x^{1-\lambda}} & -hx^{\lambda-1}\\1 & 0\end{pmatrix} \begin{pmatrix} \frac{x^{1-\lambda} + x^\lambda + h - 1}{h} & -\frac{x^{\lambda}}{h}\\ 1 & 0\end{pmatrix} \\
& \quad \quad = \frac{e^{-\gamma}}{1 - x^{1-2\lambda}} \begin{pmatrix} \frac{h^2 + (x^{1-\lambda} + (h-1) + x^\lambda)(x-x^{1-\lambda}-h)}{hx^{\lambda}} & \frac{h+x^{1-\lambda} - x}{h}\\ \frac{h^2 + (x^{1-\lambda} + (h-1) + x^\lambda)(x-x^{1-\lambda}-h)}{hx^{\lambda}} & \frac{h+x^{1-\lambda} - x}{h}\end{pmatrix}.
\end{align}
\end{widetext}
Eq.~\ref{eq:perturb3} allows us to solve for $\gamma$ which works out to
\begin{equation}
\gamma = \ln \left[\frac{x^{2(1-\lambda)} - x^{1-\lambda}(1-2h+x) - 
h + x - hx}{h(x^{1-\lambda} - x^\lambda)}\right].
\label{eq:solution}
\end{equation}
Eq.~\ref{eq:pertubativeexpansion} and Eq.~\ref{eq:solution} provide a perturbative 
solution to the heterogeneous system. 
However, the value of $\gamma$ diverges as $\lambda$ approaches $\lambda^*$, 
where $\lambda^*$ is given by 
\begin{equation}
\lambda^* = 1 - \frac{\ln \left[\frac{1 +x-2h +\sqrt{(x-1)^2+4h^2}}{2}\right]}{\ln x}
\label{eq:criticallambda}
\end{equation}
For $\lambda \geq \lambda^*$, the perturbative approach is not valid. 
Thus, in the large $N$ limit, the heterogeneous network behaves exactly like 
the translationally symmetric network up to that value $\lambda^*$, at which 
point it deviates markedly. The value $\lambda^*$ marks the location of the 
singularity in $\psi_\omega(\lambda)$ (see Fig.~\ref{fig:crossover}), which
is also the slope of the tie line in the rate function. 

The transition is accompanied by a cross-over in the behavior of the 
eigenvectors. In particular when the heterogenous network behaves like the translationally symmetric network, the largest eigenvector of $\mathbb{W}_\omega(\lambda)$ has an unbound or delocalized nature. 
In contrast, for $\lambda > \lambda^*$ the maximal eigenvector exhibits exponential localization around the heterogeneity in agreement with kinetic Monte Carlo simulations. 
The study of the triangle network, while mathematically more complicated, 
retains the same phenomenology of a dynamic phase transition between 
delocalized and localized phases (see Fig.~\ref{fig:crossover}) and the critical value of $\lambda^*$ can be determined similarly. 
Given the robustness to decorations of the ring, we expect similar localization
transitions in more general pseudo-one-dimensional cyclical networks in the limit
of a large number of states.
An analytic proof of these assertions is beyond the scope of this paper and will be provided in a subsequent publication. 
We note that the transfer matrix analysis is also amenable to a study of random 
disorder, where tuning of the disorder is known to induce a localization 
transition~\cite{Derrida1987}. 

In conclusion, we have demonstrated that singularities in trajectory space can in fact arise 
even in very simple driven cyclical kinetic networks.   
Previous demonstrations of such singularities have invariably involved
many particle systems which are not particularly amenable to exact analytical treatment. 
Dynamic phase transitions in these simple driven kinetic network systems can hence serve as toy models to elucidate the various features associated with 
phase transitions in trajectory space. 
Finally, we note that cyclical or periodic driven kinetic networks find uses in many biophysical 
contexts. 
Dynamic phase transitions in such Markov state models and possible biophysical 
implications of the consequent large deviation statistics is a subject of ongoing work and 
will be discussed in a subsequent publication.

We acknowledge useful discussions with Christopher Jarzynski and 
Fr\'{e}d\'{e}ric van Wijland. 
This work was supported in part by the Director, Office of Science, Office of Basic Energy Sciences, Materials Sciences, and Engineering Division, of the 
U.S. Department of Energy under contract No.\ DE AC02-05CH11231 
(S.V. and P. G.). 
T.G. acknowledges support from the NSF Graduate Research Fellowship and the 
Fannie and John Hertz Foundation. 
This research used the resources of the National Energy Research Scientific 
Computing Center, which was supported by the Office of Science of the US 
Department of Energy under contract No.\ DE-AC02- 05CH11231.


%\bibliography{References,References2}

\begin{thebibliography}{24}%
\makeatletter
\providecommand \@ifxundefined [1]{%
 \@ifx{#1\undefined}
}%
\providecommand \@ifnum [1]{%
 \ifnum #1\expandafter \@firstoftwo
 \else \expandafter \@secondoftwo
 \fi
}%
\providecommand \@ifx [1]{%
 \ifx #1\expandafter \@firstoftwo
 \else \expandafter \@secondoftwo
 \fi
}%
\providecommand \natexlab [1]{#1}%
\providecommand \enquote  [1]{``#1''}%
\providecommand \bibnamefont  [1]{#1}%
\providecommand \bibfnamefont [1]{#1}%
\providecommand \citenamefont [1]{#1}%
\providecommand \href@noop [0]{\@secondoftwo}%
\providecommand \href [0]{\begingroup \@sanitize@url \@href}%
\providecommand \@href[1]{\@@startlink{#1}\@@href}%
\providecommand \@@href[1]{\endgroup#1\@@endlink}%
\providecommand \@sanitize@url [0]{\catcode `\\12\catcode `\$12\catcode
  `\&12\catcode `\#12\catcode `\^12\catcode `\_12\catcode `\%12\relax}%
\providecommand \@@startlink[1]{}%
\providecommand \@@endlink[0]{}%
\providecommand \url  [0]{\begingroup\@sanitize@url \@url }%
\providecommand \@url [1]{\endgroup\@href {#1}{\urlprefix }}%
\providecommand \urlprefix  [0]{URL }%
\providecommand \Eprint [0]{\href }%
\providecommand \doibase [0]{http://dx.doi.org/}%
\providecommand \selectlanguage [0]{\@gobble}%
\providecommand \bibinfo  [0]{\@secondoftwo}%
\providecommand \bibfield  [0]{\@secondoftwo}%
\providecommand \translation [1]{[#1]}%
\providecommand \BibitemOpen [0]{}%
\providecommand \bibitemStop [0]{}%
\providecommand \bibitemNoStop [0]{.\EOS\space}%
\providecommand \EOS [0]{\spacefactor3000\relax}%
\providecommand \BibitemShut  [1]{\csname bibitem#1\endcsname}%
\let\auto@bib@innerbib\@empty
%</preamble>
\bibitem [{\citenamefont {Crooks}(2000)}]{Crooks2000}%
  \BibitemOpen
  \bibfield  {author} {\bibinfo {author} {\bibfnamefont {G.~E.}\ \bibnamefont
  {Crooks}},\ }\href@noop {} {\bibfield  {journal} {\bibinfo  {journal}
  {Physical Review E}\ }\textbf {\bibinfo {volume} {61}},\ \bibinfo {pages}
  {2361} (\bibinfo {year} {2000})}\BibitemShut {NoStop}%
\bibitem [{\citenamefont {Jarzynski}(1997)}]{Jarzynski1997}%
  \BibitemOpen
  \bibfield  {author} {\bibinfo {author} {\bibfnamefont {C.}~\bibnamefont
  {Jarzynski}},\ }\href {\doibase 10.1103/PhysRevE.56.5018} {\bibfield
  {journal} {\bibinfo  {journal} {Physical Review E}\ }\textbf {\bibinfo
  {volume} {56}},\ \bibinfo {pages} {5018} (\bibinfo {year}
  {1997})}\BibitemShut {NoStop}%
\bibitem [{\citenamefont {Jarzynski}(2011)}]{Jarzynski2011}%
  \BibitemOpen
  \bibfield  {author} {\bibinfo {author} {\bibfnamefont {C.}~\bibnamefont
  {Jarzynski}},\ }\href {\doibase 10.1146/annurev-conmatphys-062910-140506}
  {\bibfield  {journal} {\bibinfo  {journal} {Annual Review of Condensed Matter
  Physics}\ }\textbf {\bibinfo {volume} {2}},\ \bibinfo {pages} {329} (\bibinfo
  {year} {2011})}\BibitemShut {NoStop}%
\bibitem [{\citenamefont {Bodineau}\ and\ \citenamefont
  {Derrida}(2004)}]{Bodineau2004}%
  \BibitemOpen
  \bibfield  {author} {\bibinfo {author} {\bibfnamefont {T.}~\bibnamefont
  {Bodineau}}\ and\ \bibinfo {author} {\bibfnamefont {B.}~\bibnamefont
  {Derrida}},\ }\href {\doibase 10.1103/PhysRevLett.92.180601} {\bibfield
  {journal} {\bibinfo  {journal} {Physical Review Letters}\ }\textbf {\bibinfo
  {volume} {92}},\ \bibinfo {pages} {180601} (\bibinfo {year}
  {2004})}\BibitemShut {NoStop}%
\bibitem [{\citenamefont {Bodineau}\ and\ \citenamefont
  {Derrida}(2005)}]{Bodineau2005}%
  \BibitemOpen
  \bibfield  {author} {\bibinfo {author} {\bibfnamefont {T.}~\bibnamefont
  {Bodineau}}\ and\ \bibinfo {author} {\bibfnamefont {B.}~\bibnamefont
  {Derrida}},\ }\href {\doibase 10.1103/PhysRevE.72.066110} {\bibfield
  {journal} {\bibinfo  {journal} {Physical Review E}\ }\textbf {\bibinfo
  {volume} {72}},\ \bibinfo {pages} {066110} (\bibinfo {year}
  {2005})}\BibitemShut {NoStop}%
\bibitem [{\citenamefont {Espigares}\ \emph {et~al.}(2013)\citenamefont
  {Espigares}, \citenamefont {Garrido},\ and\ \citenamefont
  {Hurtado}}]{Espigares2013}%
  \BibitemOpen
  \bibfield  {author} {\bibinfo {author} {\bibfnamefont {C.~P.}\ \bibnamefont
  {Espigares}}, \bibinfo {author} {\bibfnamefont {P.~L.}\ \bibnamefont
  {Garrido}}, \ and\ \bibinfo {author} {\bibfnamefont {P.~I.}\ \bibnamefont
  {Hurtado}},\ }\href {\doibase 10.1103/PhysRevE.87.032115} {\bibfield
  {journal} {\bibinfo  {journal} {Physical Review E}\ }\textbf {\bibinfo
  {volume} {87}},\ \bibinfo {pages} {032115} (\bibinfo {year}
  {2013})}\BibitemShut {NoStop}%
\bibitem [{\citenamefont {Harris}\ \emph {et~al.}(2005)\citenamefont {Harris},
  \citenamefont {R\'{a}kos},\ and\ \citenamefont {Sch\"{u}tz}}]{Harris2005}%
  \BibitemOpen
  \bibfield  {author} {\bibinfo {author} {\bibfnamefont {R.~J.}\ \bibnamefont
  {Harris}}, \bibinfo {author} {\bibfnamefont {A.}~\bibnamefont {R\'{a}kos}}, \
  and\ \bibinfo {author} {\bibfnamefont {G.~M.}\ \bibnamefont {Sch\"{u}tz}},\
  }\href {http://stacks.iop.org/1742-5468/2005/i=08/a=P08003} {\bibfield
  {journal} {\bibinfo  {journal} {Journal of Statistical Mechanics: Theory and
  Experiment}\ }\textbf {\bibinfo {volume} {2005}},\ \bibinfo {pages} {P08003}
  (\bibinfo {year} {2005})}\BibitemShut {NoStop}%
\bibitem [{\citenamefont {Hurtado}\ and\ \citenamefont
  {Garrido}(2011)}]{Hurtado2011v2}%
  \BibitemOpen
  \bibfield  {author} {\bibinfo {author} {\bibfnamefont {P.~I.}\ \bibnamefont
  {Hurtado}}\ and\ \bibinfo {author} {\bibfnamefont {P.~L.}\ \bibnamefont
  {Garrido}},\ }\href@noop {} {\bibfield  {journal} {\bibinfo  {journal}
  {Physical Review Letters}\ }\textbf {\bibinfo {volume} {107}},\ \bibinfo
  {pages} {180601} (\bibinfo {year} {2011})}\BibitemShut {NoStop}%
\bibitem [{\citenamefont {Garrahan}\ \emph {et~al.}(2007)\citenamefont
  {Garrahan}, \citenamefont {Jack}, \citenamefont {Lecomte}, \citenamefont
  {Pitard}, \citenamefont {van Duijvendijk},\ and\ \citenamefont {van
  Wijland}}]{Garrahan2007}%
  \BibitemOpen
  \bibfield  {author} {\bibinfo {author} {\bibfnamefont {J.~P.}\ \bibnamefont
  {Garrahan}}, \bibinfo {author} {\bibfnamefont {R.~L.}\ \bibnamefont {Jack}},
  \bibinfo {author} {\bibfnamefont {V.}~\bibnamefont {Lecomte}}, \bibinfo
  {author} {\bibfnamefont {E.}~\bibnamefont {Pitard}}, \bibinfo {author}
  {\bibfnamefont {K.}~\bibnamefont {van Duijvendijk}}, \ and\ \bibinfo {author}
  {\bibfnamefont {F.}~\bibnamefont {van Wijland}},\ }\href
  {http://link.aps.org/doi/10.1103/PhysRevLett.98.195702} {\bibfield  {journal}
  {\bibinfo  {journal} {Physical Review Letters}\ }\textbf {\bibinfo {volume}
  {98}},\ \bibinfo {pages} {195702} (\bibinfo {year} {2007})}\BibitemShut
  {NoStop}%
\bibitem [{\citenamefont {Speck}\ \emph {et~al.}(2012)\citenamefont {Speck},
  \citenamefont {Engel},\ and\ \citenamefont {Seifert}}]{Speck2012v2}%
  \BibitemOpen
  \bibfield  {author} {\bibinfo {author} {\bibfnamefont {T.}~\bibnamefont
  {Speck}}, \bibinfo {author} {\bibfnamefont {A.}~\bibnamefont {Engel}}, \ and\
  \bibinfo {author} {\bibfnamefont {U.}~\bibnamefont {Seifert}},\ }\href@noop
  {} {\bibfield  {journal} {\bibinfo  {journal} {Journal of Statistical
  Mechanics: Theory and Experiment}\ }\textbf {\bibinfo {volume} {2012}},\
  \bibinfo {pages} {P12001} (\bibinfo {year} {2012})}\BibitemShut {NoStop}%
\bibitem [{\citenamefont {Bunin}\ and\ \citenamefont
  {Kafri}(2013)}]{Bunin2013}%
  \BibitemOpen
  \bibfield  {author} {\bibinfo {author} {\bibfnamefont {G.}~\bibnamefont
  {Bunin}}\ and\ \bibinfo {author} {\bibfnamefont {Y.}~\bibnamefont {Kafri}},\
  }\href {http://iopscience.iop.org/1751-8121/46/9/095002} {\bibfield
  {journal} {\bibinfo  {journal} {Journal of Physics A: Mathematical and
  General}\ }\textbf {\bibinfo {volume} {46}},\ \bibinfo {pages} {1} (\bibinfo
  {year} {2013})}\BibitemShut {NoStop}%
\bibitem [{\citenamefont {Kolomeisky}\ and\ \citenamefont
  {Fisher}(2007)}]{Kolomeisky2007}%
  \BibitemOpen
  \bibfield  {author} {\bibinfo {author} {\bibfnamefont {A.~B.}\ \bibnamefont
  {Kolomeisky}}\ and\ \bibinfo {author} {\bibfnamefont {M.~E.}\ \bibnamefont
  {Fisher}},\ }\href {\doibase 10.1146/annurev.physchem.58.032806.104532}
  {\bibfield  {journal} {\bibinfo  {journal} {Annual Review of Physical
  Chemistry}\ }\textbf {\bibinfo {volume} {58}},\ \bibinfo {pages} {675}
  (\bibinfo {year} {2007})}\BibitemShut {NoStop}%
\bibitem [{\citenamefont {Fisher}\ and\ \citenamefont
  {Kolomeisky}(1999)}]{Fisher1999}%
  \BibitemOpen
  \bibfield  {author} {\bibinfo {author} {\bibfnamefont {M.~E.}\ \bibnamefont
  {Fisher}}\ and\ \bibinfo {author} {\bibfnamefont {A.~B.}\ \bibnamefont
  {Kolomeisky}},\ }\href {\doibase 10.1073/pnas.96.12.6597} {\bibfield
  {journal} {\bibinfo  {journal} {Proceedings of the National Academy of
  Sciences}\ }\textbf {\bibinfo {volume} {96}},\ \bibinfo {pages} {6597}
  (\bibinfo {year} {1999})},\ \Eprint
  {http://arxiv.org/abs/http://www.pnas.org/content/96/12/6597.full.pdf+html}
  {http://www.pnas.org/content/96/12/6597.full.pdf+html} \BibitemShut {NoStop}%
\bibitem [{\citenamefont {Tu}(2008)}]{Tu2008}%
  \BibitemOpen
  \bibfield  {author} {\bibinfo {author} {\bibfnamefont {Y.}~\bibnamefont
  {Tu}},\ }\href {\doibase 10.1073/pnas.0804641105} {\bibfield  {journal}
  {\bibinfo  {journal} {Proceedings of the National Academy of Sciences of the
  United States of America}\ }\textbf {\bibinfo {volume} {105}},\ \bibinfo
  {pages} {11737} (\bibinfo {year} {2008})}\BibitemShut {NoStop}%
\bibitem [{\citenamefont {Hopfield}(1974)}]{Hopfield1974}%
  \BibitemOpen
  \bibfield  {author} {\bibinfo {author} {\bibfnamefont {J.~J.}\ \bibnamefont
  {Hopfield}},\ }\href {http://www.pnas.org/content/71/10/4135.abstract}
  {\bibfield  {journal} {\bibinfo  {journal} {Proceedings of the National
  Academy of Sciences}\ }\textbf {\bibinfo {volume} {71}},\ \bibinfo {pages}
  {4135} (\bibinfo {year} {1974})}\BibitemShut {NoStop}%
\bibitem [{\citenamefont {Murugan}\ \emph {et~al.}(2012)\citenamefont
  {Murugan}, \citenamefont {Huse},\ and\ \citenamefont
  {Leibler}}]{Murugan2012v2}%
  \BibitemOpen
  \bibfield  {author} {\bibinfo {author} {\bibfnamefont {A.}~\bibnamefont
  {Murugan}}, \bibinfo {author} {\bibfnamefont {D.~A.}\ \bibnamefont {Huse}}, \
  and\ \bibinfo {author} {\bibfnamefont {S.}~\bibnamefont {Leibler}},\
  }\href@noop {} {\bibfield  {journal} {\bibinfo  {journal} {Proceedings of the
  National Academy of Sciences}\ }\textbf {\bibinfo {volume} {109}},\ \bibinfo
  {pages} {12034} (\bibinfo {year} {2012})}\BibitemShut {NoStop}%
\bibitem [{Note1()}]{Note1}%
  \BibitemOpen
  \bibinfo {note} {Such motifs were introduced, for example, in some of the
  earliest models of kinetic proofreading~\cite
  {Hopfield1974,Bennett1979}}\BibitemShut {NoStop}%
\bibitem [{\citenamefont {Seifert}(2012)}]{Seifert2012}%
  \BibitemOpen
  \bibfield  {author} {\bibinfo {author} {\bibfnamefont {U.}~\bibnamefont
  {Seifert}},\ }\href {http://stacks.iop.org/0034-4885/75/i=12/a=126001}
  {\bibfield  {journal} {\bibinfo  {journal} {Reports on Progress in Physics}\
  }\textbf {\bibinfo {volume} {75}},\ \bibinfo {pages} {126001} (\bibinfo
  {year} {2012})}\BibitemShut {NoStop}%
\bibitem [{\citenamefont {Gillespie}(1976)}]{Gillespie1976}%
  \BibitemOpen
  \bibfield  {author} {\bibinfo {author} {\bibfnamefont {D.~T.}\ \bibnamefont
  {Gillespie}},\ }\href@noop {} {\bibfield  {journal} {\bibinfo  {journal}
  {Journal of Computational Physics}\ }\textbf {\bibinfo {volume} {22}},\
  \bibinfo {pages} {403} (\bibinfo {year} {1976})}\BibitemShut {NoStop}%
\bibitem [{\citenamefont {Touchette}(2009)}]{Touchette2009}%
  \BibitemOpen
  \bibfield  {author} {\bibinfo {author} {\bibfnamefont {H.}~\bibnamefont
  {Touchette}},\ }\href {\doibase
  http://dx.doi.org/10.1016/j.physrep.2009.05.002} {\bibfield  {journal}
  {\bibinfo  {journal} {Physics Reports}\ }\textbf {\bibinfo {volume} {478}},\
  \bibinfo {pages} {1} (\bibinfo {year} {2009})}\BibitemShut {NoStop}%
\bibitem [{\citenamefont {Lebowitz}\ and\ \citenamefont
  {Spohn}(1999)}]{Lebowitz1999}%
  \BibitemOpen
  \bibfield  {author} {\bibinfo {author} {\bibfnamefont {J.}~\bibnamefont
  {Lebowitz}}\ and\ \bibinfo {author} {\bibfnamefont {H.}~\bibnamefont
  {Spohn}},\ }\href
  {http://link.springer.com/article/10.1023/A\%3A1004589714161} {\bibfield
  {journal} {\bibinfo  {journal} {Journal of Statistical Physics}\ }\textbf
  {\bibinfo {volume} {95}},\ \bibinfo {pages} {333} (\bibinfo {year}
  {1999})}\BibitemShut {NoStop}%
\bibitem [{\citenamefont {Touchette}\ and\ \citenamefont
  {Harris}(2013)}]{Touchette2013}%
  \BibitemOpen
  \bibfield  {author} {\bibinfo {author} {\bibfnamefont {H.}~\bibnamefont
  {Touchette}}\ and\ \bibinfo {author} {\bibfnamefont {R.~J.}\ \bibnamefont
  {Harris}},\ }\href@noop {} {\bibfield  {journal} {\bibinfo  {journal}
  {Nonequilibrium Statistical Physics of Small Systems: Fluctuation Relations
  and Beyond}\ ,\ \bibinfo {pages} {335}} (\bibinfo {year} {2013})}\BibitemShut
  {NoStop}%
\bibitem [{\citenamefont {Derrida}\ \emph {et~al.}(1987)\citenamefont
  {Derrida}, \citenamefont {Mecheri},\ and\ \citenamefont
  {Pichard}}]{Derrida1987}%
  \BibitemOpen
  \bibfield  {author} {\bibinfo {author} {\bibfnamefont {B.}~\bibnamefont
  {Derrida}}, \bibinfo {author} {\bibfnamefont {K.}~\bibnamefont {Mecheri}}, \
  and\ \bibinfo {author} {\bibfnamefont {J.~L.}\ \bibnamefont {Pichard}},\
  }\href {http://dx.doi.org/10.1051/jphys:01987004805073300} {\bibfield
  {journal} {\bibinfo  {journal} {J. Phys. France}\ }\textbf {\bibinfo {volume}
  {48}},\ \bibinfo {pages} {733} (\bibinfo {year} {1987})}\BibitemShut
  {NoStop}%
\bibitem [{\citenamefont {Bennett}(1979)}]{Bennett1979}%
  \BibitemOpen
  \bibfield  {author} {\bibinfo {author} {\bibfnamefont {C.~H.}\ \bibnamefont
  {Bennett}},\ }\href@noop {} {\bibfield  {journal} {\bibinfo  {journal}
  {BioSystems}\ }\textbf {\bibinfo {volume} {11}},\ \bibinfo {pages} {85}
  (\bibinfo {year} {1979})}\BibitemShut {NoStop}%
\end{thebibliography}
\end{document}